\documentclass[12pt]{article}%
\usepackage{amsfonts}
\usepackage{graphicx}
\usepackage{amsmath}
\usepackage{amssymb}%
\pdfoutput=1
\makeatletter

\@addtoreset{equation}{section}
\makeatother
\newcommand{\beq}{\begin{equation}}
\newcommand{\eeq}{\end{equation}}

\begin{document}
\baselineskip=18pt
\baselineskip 0.7cm

\begin{titlepage}
\setcounter{page}{0}
\renewcommand{\thefootnote}{\fnsymbol{footnote}}
\begin{flushright}
\end{flushright}
\vskip 1.5cm
\begin{center}
{\LARGE \bf
The Hodge Operator Revisited
\vskip 1.5cm
}
{\large
L. Castellani$^{~a,b,}$\footnote{leonardo.castellani@uniupo.it},
R. Catenacci$^{~a,c,}$\footnote{roberto.catenacci@uniupo.it},
and
P.A. Grassi$^{~a,b,}$\footnote{pietro.grassi@uniupo.it}
\medskip
}
\vskip 0.5cm
{
\small\it
\centerline{$^{(a)}$ Dipartimento di Scienze e Innovazione Tecnologica, Universit\`a del Piemonte Orientale} }
\centerline{\it Viale T. Michel, 11, 15121 Alessandria, Italy}
\medskip
\centerline{$^{(b)}$ {\it
INFN, Sezione di Torino, via P. Giuria 1, 10125 Torino} }
\centerline{$^{(c)}$ {\it
Gruppo Nazionale di Fisica Matematica, INdAM, P.le Aldo Moro 5, 00185 Roma} }
\vskip  .5cm
\medskip
\end{center}
\centerline{{\bf Abstract}}
\medskip
\noindent
We present a new construction for the Hodge operator for differential manifolds
based on a Fourier (Berezin)-integral representation. We find a simple formula for the Hodge
dual of the wedge product of differential forms, using the (Berezin)-convolution.
The present analysis is easily extended to supergeometry and to non-commutative geometry.
\bigskip
\bigskip
\noindent
\bigskip

\end{titlepage}
\setcounter{page}{1}

\vfill
\eject

\newpage\setcounter{footnote}{0} \newpage\setcounter{footnote}{0}

\section{Introduction}

Pursuing the construction of supersymmetric Lagrangians based in the framework
of supermanifold geometry, we proposed in \cite{Castellani:2015paa} a new
\textit{Hodge operator} $\star$ acting on (super)differential forms. For that
aim, we have developed a complete formalism (integral-, pseudo- e super-forms,
their complexes and the integration theory) in a series of papers
\cite{Castellani:2014goa,Castellani:2015paa,Castellani:2015ata} together with
a suitable Hodge operator. As a byproduct, this new mathematical tool sheds
also a new light on the Hodge operator in conventional differential geometry.
The present work illustrates this new point of view.

The Hodge operator plays an essential role in differential geometry, yielding
a fundamental relation between the exterior bundle $\Omega^{\bullet}(M)$ of
differential forms and the scalar product $(\bullet,\bullet)$ on the manifold.
The construction requires the existence of a metric $g$ on the the manifold
$\mathcal{M}$ and is an involutive operation $\star$ which satisfies the
linearity condition $\star(f\omega)=f\star\omega$ with $\omega$ a given
$p$-form of $\Omega^{\bullet}(M)$.

In the case of supermanifolds (we refer for ex. to \cite{Castellani:2015paa}
for the basic ingredients of supergeometry; see also \cite{Voronov2} and for a
recent extensive review see \cite{Witten:2012bg}), the definition of the Hodge
dual turns out to be harder than expected since one has to deal with the
infinite-dimensional complexes of superforms. The integral forms and
pseudo-forms are crucial to establish the correct matching of elements between
the different spaces of forms. This new type of differential forms requires
the enlargement of the conventional space spanned by the fundamental
$1$-forms, admitting distribution-like expressions (essentially, Dirac delta
functions and Heaviside step functions). This has triggered us to consider the
\textit{Fourier analysis} for differential forms (this was also considered in
\cite{Kalkman:1993zp}), and leads to an integral Fourier representation of the
Hodge operator as explained in \cite{Castellani:2015paa,Castellani:2015ata}.
Such a representation can first be established in the case of a conventional
manifold $M$ without any reference to supermanifolds, except for the
notion of \textit{Berezin integral}. A new set of anticommuting variables
playing the role of dual variables to fundamental 1-forms $dx^{i}$ are
introduced and the Hodge operator is defined by a suitable Berezin integration
on the new variables. The result is proven to coincide with the usual Hodge
operator. When extended to supermanifolds, our construction yields a ``good"
definition of Hodge operator, satisfying all desired properties.

As is well known, in conventional Fourier analysis, the Fourier transform of a
product of two functions is the \textit{convolution integral} of the Fourier
transform of the two functions. In a reciprocal way, the Fourier transform of
the convolution integral of two functions can be expressed as the product of
the Fourier transforms of the latter. This simple formula can be imported in
our framework where the Fourier transform represents the Hodge operator and
the convolution integral is a suitable Berezin integral of two differential
forms. With this observation we are able to express the Hodge dual of the
wedge product of two differential forms as the (Berezin)-convolution of the
Hodge duals of the differential forms. In the present paper, we show the
consistency of these definitions, derive their properties and make contact
with the conventional results.

Finally, our Fourier (Berezin) integral representation of the Hodge dual
operator can be extended to noncommutative spaces. A very recent work on this
appeared in \cite{Majid}.

The letter is organized as follows: in Sec. 2, we review the integration
theory of differential forms as Berezin integration. In Sec. 3 we give the
integral Fourier representation of the Hodge dual. In Sec. 4, we present the
Berezin-convolution and the Hodge dual of the product of differential forms.

\section{Forms and Integration}

The usual integration theory of differential forms for bosonic manifolds can
be conveniently rephrased to uncover its relation with Berezin integration.

We start with a simple example: consider in $\mathbb{R}$ the integrable 1-form
$\omega$ $=g(x)dx$ (with $g(x)$ an integrable function in $\mathbb{R}$ ). We
have:
\[
\int_{\mathbb{R}}\omega=\int_{\mathbb{-\infty}}^{+\infty}g(x)dx\,.
\]
Observing that $dx$ is an anticommuting quantity, and denoting it by $\psi$,
we could think of $\omega$ as a function on the superspace $\mathbb{R}^{1|1}$:%
\begin{equation}
\omega=g(x)dx=f(x,\psi)=g(x)\psi
\end{equation}
This function can be integrated \textit{\`{a} la} Berezin reproducing the
usual definition:%
\[
\int_{\mathbb{R}^{1|1}}f(x,\psi)[dxd\psi]=\int_{\mathbb{-\infty}}^{+\infty
}g(x)dx\,=\int_{\mathbb{R}}\omega
\]
Note that the symbol of the formal measure $[dxd\psi]$ is written just to
emphasize that we are integrating on the \textbf{two} variables $x$ and $\psi
$, hence the $dx$ inside $[dxd\psi]$ is \textit{not} identified with $\psi$.

Denoting by $M$ a bosonic orientable differentiable manifold of dimension $n$,
its exterior bundle $\Omega^{\bullet}(M)=\sum_{p=0}^{n}\bigwedge^{p}(M)$ is
the direct sum of $\bigwedge^{p}(M)$ (sometimes denoted also by $\Omega
^{p}(M)$). A section $\omega$ of $\Omega^{p}(M)$ can be written locally as
\begin{equation}
\omega=\sum\omega_{i_{1}\dots i_{p}}(x)dx^{i_{1}}\wedge\dots\wedge dx^{i_{p}}
\label{inA}%
\end{equation}
where the coefficients $\omega_{i_{1}\dots i_{p}}(x)$ are functions on $M$ and
$i_{1}<...<i_{p}$. The integral of $\omega\in\Omega^{n}(M)$ is defined as:
\begin{equation}
I[\omega]=\int_{M}\omega=\int_{M}\omega_{i_{1}\dots i_{n}}(x)\,d^{n}x\,,
\label{inB}%
\end{equation}
This opens the way to relating the integration theory of forms and the Berezin
integral, by substituting every $1$-form $dx^{i}$ with a corresponding
abstract Grassmann variable denoted again with $dx^{i}.$ A section $\omega$ of
${\Omega^{\bullet}(}M{)}$ is viewed locally as a function on a supermanifold
$\mathcal{M=}T^{\ast}(M)$ with coordinates $(x^{i},dx^{i}):$
\begin{equation}
\omega(x,dx)=\sum\omega_{i_{1}\dots i_{p}}(x)dx^{i_{1}}\dots dx^{i_{p}}\,;
\label{inAA}%
\end{equation}
such functions are polynomials in $dx$'s. Supposing now that the form $\omega$
is integrable, its Berezin integral gives:
\begin{equation}
\int_{\mathcal{M=}T^{\ast}(M))}\omega(x,dx)[d^{n}xd^{n}(dx)]=\int_{M}\omega
\end{equation}

\section{The Integral Representation of the Hodge Star}

In the following, for a given set $\{\xi^{i}\}_{i=1}^{n}$ of Grassmann
variables, our definition of the Berezin integral is $\int_{\mathbb{R}^{0|n}%
}\xi^{1}...\xi^{n}\left[  d^{n}\xi\right]  =1$ and not $\int_{\mathbb{R}%
^{0|n}}\xi^{1}...\xi^{n}\left[  d^{n}\xi\right]  =\left(  -1\right)
^{\frac{n(n-1)}{2}}.$ Moreover, if $\alpha$ is a monomial expression of some
anticommuting variables $\alpha^{k}$ not depending on the $\xi^{i},$ we
define: $\int_{\mathbb{R}^{0|n}}\alpha\xi^{1}...\xi^{n}\left[  d^{n}%
\xi\right]  =\alpha,$ where the product between $\alpha$ and the $\xi^{i}$ is
the usual $\mathbb{Z}_{2}$ graded wedge product in the superalgebra generated
by the tensor product of the Grassmann algebra generated by the $\xi^{i}$ and
that generated by the $\alpha^{k}:$ if $\mathcal{A}$ and $\mathcal{B}$ are two
$\mathbb{Z}_{2}$-graded algebras with products $\cdot_{\mathcal{A}}$and
$\cdot_{\mathcal{B}}$, the $\mathbb{Z}_{2}$-graded tensor product
$\mathcal{A}\otimes\mathcal{B}$ is a $\mathbb{Z}_{2}$-graded algebra with the
product (for homogeneous elements) given by :%
\[
(a\otimes b)\cdot_{\mathcal{A}\otimes\mathcal{B}}(a^{\prime}\otimes b^{\prime
})=(-1)^{\left\vert a^{\prime}\right\vert \left\vert b\right\vert }%
a\cdot_{\mathcal{A}}a^{\prime}\otimes b\cdot_{\mathcal{B}}b^{\prime}%
\]
In our case the algebras are Grassmann algebras and the products $\cdot$ are
wedge products. The symbols $\otimes$ and $\wedge$ will be, in general, omitted.

As observed in \cite{Castellani:2015paa} one can obtain the usual Hodge dual
in $\mathbb{R}^{n}$\textbf{ }(for a metric given by a matrix $A$ with entries
$g_{ij}$ ) by means of\textbf{ }the Fourier (Berezin)-integral transform $\mathcal{T}$ . For
$\omega(x,dx)\in\Omega^{k}(\mathbb{R}^{n})$ we have:
\begin{equation}
\star\omega=i^{\left(  k^{2}-n^{2}\right)  }\frac{\sqrt{\left\vert
g\right\vert }}{g}\mathcal{T}(\omega)=i^{\left(  k^{2}-n^{2}\right)  }%
\frac{\sqrt{\left\vert g\right\vert }}{g}\int_{\mathbb{R}^{0|n}}\omega
(x,\eta^{\prime})e^{idxA\eta^{\prime}}[d^{n}\eta^{\prime}] \label{duale1}%
\end{equation}
where $g=\mathrm{det}A$ and the exponential series defining $e^{idxA\eta
^{\prime}}$ is written using the $\mathbb{Z}_{2}$ graded wedge product quoted
above. The Grassmann variables $\eta^{\prime}$ are defined as $\eta^{\prime
}=A^{-1}\eta$ where the $\eta$ are the \textit{(parity changed)} variables
dual to the $dx$. In this way the covariance properties of $\omega
(x,\eta^{\prime})$ are exactly those of a differential form.

For example, in $\mathbb{R}^{2}$\textbf{ }we can compute:%
\begin{equation}
e^{idxA\eta^{\prime}}=1+ig_{11}dx^{1}\eta^{\prime1}+ig_{21}dx^{2}\eta
^{\prime1}+ig_{12}dx^{1}\eta^{\prime2}+ig_{22}dx^{2}\eta^{\prime2}%
+gdx^{1}dx^{2}\eta^{\prime1}\eta^{\prime2} \label{fourier5}%
\end{equation}
and the definition (\ref{duale1}) gives the usual results:
\begin{align*}
\star1  &  =i^{\left(  0^{2}-2^{2}\right)  }\mathcal{T}(1)=\frac
{\sqrt{\left\vert g\right\vert }}{g}\int_{\mathbb{R}^{0|2}}e^{idxA\eta
^{\prime}}[d^{2}\eta^{\prime}]=\sqrt{\left\vert g\right\vert }dx^{1}dx^{2}\\
\star dx^{1}dx^{2}  &  =i^{\left(  2^{2}-2^{2}\right)  }\mathcal{T}%
(\eta^{\prime1}\eta^{\prime2})=\frac{\sqrt{\left\vert g\right\vert }}{g}%
\int_{\mathbb{R}^{0|2}}\eta^{\prime1}\eta^{\prime2}e^{idxA\eta}[d^{2}%
\eta^{\prime}]=\frac{\sqrt{\left\vert g\right\vert }}{g}\\
\star dx^{1}  &  =i^{\left(  1^{2}-2^{2}\right)  }\mathcal{T}(\eta^{\prime
1})=i^{\left(  1^{2}-2^{2}\right)  }\frac{\sqrt{\left\vert g\right\vert }}%
{g}\int_{\mathbb{R}^{0|2}}\eta^{\prime1}e^{idxA\eta^{\prime}}[d^{2}%
\eta^{\prime}]=-g^{12}\sqrt{\left\vert g\right\vert }dx^{1}+g^{11}%
\sqrt{\left\vert g\right\vert }dx^{2}\\
\star dx^{2}  &  =i^{\left(  1^{2}-2^{2}\right)  }\mathcal{T}(\eta^{\prime
2})=i^{\left(  1^{2}-2^{2}\right)  }\frac{\sqrt{\left\vert g\right\vert }}%
{g}\int_{\mathbb{R}^{0|2}}\eta^{\prime2}e^{idxA\eta^{\prime}}[d^{2}%
\eta^{\prime}]=-g^{22}\sqrt{\left\vert g\right\vert }dx^{1}+g^{21}%
\sqrt{\left\vert g\right\vert }dx^{2}%
\end{align*}
The factor $i^{\left(  k^{2}-n^{2}\right)  }$ can be obtained by computing the
transformation of the monomial form $dx^{1}dx^{2}...dx^{k}\ $in the simple
case $A=I.$

Noting that in the Berezin integral only the higher degree term in the $\eta$
variables is involved, and that the monomials $dx^{i}\eta^{i}$ are even
objects, we find:%
\begin{align*}
\mathcal{T}\left(  dx^{1}...dx^{k}\right)   &  =\int_{\mathbb{R}^{0|n}}%
\eta^{1}...\eta^{k}e^{idx\eta}[d^{n}\eta]=\\
&  =\int_{\mathbb{R}^{0|n}}\eta^{1}...\eta^{k}e^{i\left(  \sum_{i=1}^{k}%
dx^{i}\eta^{i}+\sum_{i=k+1}^{n}dx^{i}\eta^{i}\right)  }[d^{n}\eta]=\\
&  =\int_{\mathbb{R}^{0|n}}\eta^{1}...\eta^{k}e^{i\sum_{i=1}^{k}dx^{i}\eta
^{i}}e^{i\sum_{i=k+1}^{n}dx^{i}\eta^{i}}[d^{n}\eta]=\\
&  =\int_{\mathbb{R}^{0|n}}\eta^{1}...\eta^{k}e^{i\sum_{i=k+1}^{n}dx^{i}%
\eta^{i}}[d^{n}\eta]=\\
&  =\int_{\mathbb{R}^{0|n}}\frac{i^{n-k}}{\left(  n-k\right)  !}\eta
^{1}...\eta^{k}\left(  \sum_{i=k+1}^{n}dx^{i}\eta^{i}\right)  ^{n-k}[d^{n}%
\eta]
\end{align*}
Rearranging the monomials $dx^{i}\eta^{i}$ one obtains:
\[
\left(  \sum_{i=k+1}^{n}dx^{i}\eta^{i}\right)  ^{n-k}=\left(  n-k\right)
!\left(  dx^{k+1}\eta^{k+1})(dx^{k+2}\eta^{k+2})...(dx^{n}\eta^{n}\right)  =
\]%
\[
=\left(  n-k\right)  !(-1)^{\frac{1}{2}(n-k)(n-k-1)}\left(  dx^{k+1}%
dx^{k+2}...dx^{n})(\eta^{k+1}\eta^{k+2}...\eta^{n}\right)
\]
and finally:%
\[
\mathcal{T}\left(  dx^{1}...dx^{k}\right)  =
\]%
\[
=\int_{\mathbb{R}^{0|n}}\frac{i^{n-k}}{\left(  n-k\right)  !}\eta^{1}%
...\eta^{k}\left(  n-k\right)  !(-1)^{\frac{1}{2}(n-k)(n-k-1)}\left(
dx^{k+1}dx^{k+2}...dx^{n})(\eta^{k+1}\eta^{k+2}...\eta^{n}\right)  [d^{n}%
\eta]=
\]%
\[
=\int_{\mathbb{R}^{0|n}}i^{n-k}(-1)^{\frac{1}{2}(n-k)(n-k-1)}(-1)^{k(n-k)}%
\left(  dx^{k+1}dx^{k+2}...dx^{n})(\eta^{1}...\eta^{k})(\eta^{k+1}\eta
^{k+2}...\eta^{n}\right)  [d^{n}\eta]=
\]%
\[
=i^{\left(  n^{2}-k^{2}\right)  }(dx^{k+1}dx^{k+2}...dx^{n})
\]
The computation above gives immediately:%
\begin{equation}
i^{\left(  k^{2}-n^{2}\right)  }\mathcal{T}\left(  dx^{1}...dx^{k}\right)
=\star\left(  dx^{1}...dx^{k}\right)
\end{equation}
and%
\begin{equation}
\mathcal{T}^{2}\left(  \omega\right)  =i^{\left(  n^{2}-k^{2}\right)
}i^{\left(  k^{2}\right)  }\left(  \omega\right)  =i^{n^{2}}\left(
\omega\right)  \label{quadratodifourier}%
\end{equation}
yielding the usual relation:%
\begin{equation}
\star\star\omega=i^{(\left(  n-k)^{2}-n^{2}\right)  }i^{\left(  k^{2}%
-n^{2}\right)  }i^{n^{2}}(\omega)=(-1)^{k(k-n)}(\omega)
\end{equation}

\section{Convolution Product of Forms}

As for functions, one can define a convolution product between differential
forms on an ordinary manifold. The starting point is the interpretation of
differential forms as functions of the commuting variables $x$ and the
anticommuting variables $dx.$ For $\alpha\in\Omega^{p}(\mathbb{R}^{n})$ and
$\beta\in\Omega^{q}(\mathbb{R}^{n}),$ the convolution product $\bullet$ is
defined using Berezin integration on the anticommuting variables:%
\begin{equation}
\alpha\bullet\beta(x,dx)=\int_{\mathbb{R}^{0|n}}\alpha(x,\xi)\beta
(x,dx-\xi)[d^{n}\xi]\label{convoluzione}%
\end{equation}
where $\xi$ is an auxiliary anticommuting variable. The convolution product
maps $\Omega^{p}\times\Omega^{q}\rightarrow\Omega^{p+q-n}$ \footnote{We must
integrate generically monomials of the type $\left(  \xi\right)
^{p+q-k}\left(  dx\right)  ^{k}$ and the Berezin integration selects
$k=p+q-n.$}. To obtain (generically) non trivial results we must have $0\leq
p+q-n\leq n.$ The algebra of this convolution is
\[
\alpha\bullet\beta=(-1)^{\left(  n^{2}+pq\right)  }\beta\bullet\alpha
\]
The convolution `interacts" well with the integral transformation
$\mathcal{T}$ defined above and the wedge product. We will consider explicitly
only the standard bosonic case in which the matrix $A$ of the.previous
paragraph is the identity matrix $I.$

For example, in the case $n=4,$ we can compute $\mathcal{T}\left(
dx^{1}dx^{2}\right)  =dx^{3}dx^{4}$ and $\mathcal{T}\left(  dx^{1}\right)
=\left(  -i\right)  dx^{2}dx^{3}dx^{4},$ $\mathcal{T}\left(  dx^{2}\right)
=idx^{1}dx^{3}dx^{4}.$ The convolution is:%
\[
\mathcal{T}\left(  dx^{1}\right)  \bullet\mathcal{T}\left(  dx^{2}\right)
=\int_{\mathbb{R}^{0|4}}\left(  -i\right)  \xi^{2}\xi^{3}\xi^{4}\left(
i\right)  \left(  dx^{1}-\xi^{1}\right)  \left(  dx^{3}-\xi^{3}\right)
\left(  dx^{4}-\xi^{4}\right)  [d^{4}\xi]
\]
\[
=dx^{3}dx^{4}=\mathcal{T}\left(  dx^{1}dx^{2}\right)
\]
Another simple example is the case $q=n-p$ where we find:%
\begin{equation}
i^{n^{2}}\left(  -1\right)  ^{p}\left(  -1\right)  ^{p\left(  n-p\right)
}\mathcal{T}\left(  \alpha\beta\right)  =\mathcal{T}\left(  \alpha\right)
\bullet\mathcal{T}\left(  \beta\right)  \label{convoluzione1}%
\end{equation}

\noindent Indeed, recalling that:
\begin{equation}
\mathcal{T}\left(  dx^{1}...dx^{p}\right)  =i^{\left(  n^{2}-p^{2}\right)
}(dx^{p+1}dx^{p+2}...dx^{n})
\end{equation}%
\begin{equation}
\mathcal{T}\left(  dx^{p+1}...dx^{n}\right)  =i^{\left(  p^{2}\right)
}(dx^{1}dx^{2}...dx^{p})
\end{equation}%
\begin{equation}
\mathcal{T}\left(  dx^{1}...dx^{n}\right)  =1
\end{equation}
we find:%
\begin{align}
\mathcal{T}\left(  dx^{1}...dx^{p}\right)  \bullet\mathcal{T}\left(
dx^{p+1}...dx^{n}\right)   &  =i^{n^{2}}\int_{\mathbb{R}^{0|n}}(\xi
^{p+1}...\xi^{n})(dx^{1}-\xi^{1})...(dx^{p}-\xi^{p})\left[  d^{n}\xi\right]
=\\
i^{n^{2}}\int_{\mathbb{R}^{0|n}}(\xi^{p+1}...\xi^{n})\left(  -1\right)
^{p}\xi^{1}...\xi^{p}\left[  d^{n}\xi\right]   &  =i^{n^{2}}\left(  -1\right)
^{p}\left(  -1\right)  ^{p\left(  n-p\right)  }\mathcal{T}\left(  dx^{1}%
dx^{2}...dx^{n}\right)
\end{align}
\textbf{ }The properties of the convolution reflect on corresponding
properties of the Hodge star operator. Using $\star\omega=i^{\left(  k^{2}-n^{2}\right)  }\mathcal{T}(\omega)$ for $\omega(x,dx)\in\Omega^{k}(\mathbb{R}^{n})$,
 we obtain a
simple formula for the Hodge dual of the wedge product of forms in the case
$p+q=n$:%
\begin{equation}
\star\left(  \alpha\beta\right)  =\left(  -1\right)  ^{p}\left(  \star
\alpha\right)  \bullet\left(  \star\beta\right)  \label{convoluzionehodge}%
\end{equation}

Considering now the general case of a $p$-form $\alpha$ and a $q$-form $\beta$
in a $n$-dimensional space, one can prove the following relation:
\begin{equation}
\star\left(  \alpha\beta\right)  =\left(  -1\right)  ^{n+q(n-p)}\left(
\star\alpha\right)  \bullet\left(  \star\beta\right)
\label{convoluzionehodgegeneral}%
\end{equation}
\noindent easily checked to be satisfied by the monomials
\begin{equation}
\alpha=dx^{1}dx^{2}...dx^{p},~~~\beta=dx^{n-q+1}dx^{n-q+2}...dx^{n}%
\end{equation}
\noindent Indeed recall that
\begin{equation}
\star\alpha=dx^{p+1}...dx^{n},~~~\star\beta=(-1)^{q(n-q)}dx^{1}...dx^{n-q}%
\end{equation}%
\begin{equation}
\star(\alpha\beta)=(-1)^{q(n-p-q)}dx^{p+1}...dx^{n-q}%
\end{equation}
\noindent Moreover, using the definition of the convolution, one finds
\begin{equation}
(\star\alpha)\bullet(\star\beta)=(-1)^{q(n-q)}(-1)^{p}(-1)^{n(n-q-p)}%
(-1)^{p(n-p)}dx^{p+1}...dx^{n-q}%
\end{equation}
\noindent Comparing the last two equations, relation
(\ref{convoluzionehodgegeneral}) follows. By linearity the same relation
(\ref{convoluzionehodgegeneral}) holds also for generic forms. Two particular
cases provide nontrivial checks:

i) when $\alpha=1\in\Omega^{0}$ :%
\begin{align}
\star\left(  1\beta\right)  =\left(  -1\right)  ^{n+qn}\left(  \star1\right)
\bullet\left(  \star\beta\right)  =\left(  -1\right)  ^{n+qn}\int
_{\mathbb{R}^{0|n}}(\xi^{1}...\xi^{n})\left(  \star\beta(dx-\xi)\right)
\left[  d^{n}\xi\right] \nonumber\\
=\left(  -1\right)  ^{n+qn}\left(  -1\right)  ^{n(n-q)}\star\beta=\star\beta
\end{align}

ii) when $\beta=1\in\Omega^{0}$:%
\begin{align}
\star\left(  \alpha1\right)  =\left(  -1\right)  ^{n}\left(  \star
\alpha\right)  \bullet\left(  \star1\right)  =\left(  -1\right)  ^{n}%
\int_{\mathbb{R}^{0|n}}\left(  \star\alpha\right)  \left(  \xi\right)
(dx^{1}-\xi^{1})...(dx^{n}-\xi^{n})\left[  d^{n}\xi\right] \nonumber\\
=\left(  -1\right)  ^{n}\int_{\mathbb{R}^{0|n}}\left(  \star\alpha\right)
\left(  \xi+dx\right)  \left(  -1\right)  ^{n}\xi^{1}...\xi^{n}\left[
d^{n}\xi\right]  =\star\alpha
\end{align}
where we used the traslational invariance (under $\xi\rightarrow\xi+dx$) of
the Berezin integral.

Similar relations hold (modulo some suitable multiplicative coefficient
depending also on the metric) for the more general integral transform that
gives the Hodge dual for a generic metric $A.$

The convolution defined in the formula (\ref{convoluzione}) could be
normalized as:%
\begin{equation}
\alpha\bullet'\beta(x,dx)=(-1)^{\left(  n+pn+pq\right)  } \alpha \bullet
\beta
\end{equation}
Where again $p$ is the degree of $\alpha,$ $q$ the degree of $\beta,$ and $n$
the dimension of the space.

With this normalization the formula (\ref{convoluzionehodgegeneral}) looks
better:
\begin{equation}
\star\left(  \alpha\beta\right)  =\left(  \star\alpha\right)  \bullet'\left(
\star\beta\right)
\end{equation}
Indeed, noting that $\left(  \star\alpha\right)  \bullet\left(
\star\beta\right)=\left(-1\right)^{n+(n-p)n+(n-p)(n-q)}\left(  \star\alpha\right)  \bullet'\left(
\star\beta\right)$, we have:
\[
\star\left(  \alpha\beta\right) =
\left(  -1\right)  ^{n+q(n-p)}\left(-1\right)^{n+(n-p)n+(n-p)(n-q)}\left(  \star\alpha\right)  \bullet'\left(
\star\beta\right)=\left(  \star\alpha\right)  \bullet'\left(
\star\beta\right)
\]
The algebra of this new convolution is:
\begin{equation}
\alpha\bullet'\beta=(-1)^{\left(  n-p\right)  \left(  n-q\right)  }\beta
\bullet'\alpha
\end{equation}

\noindent Clearly this normalized convolution product has a unit, the standard volume form
$\star1.$

As last remark, we point out that, using our Fourier representation of the
Hodge dual, it is easy to deduce the standard formula
\begin{equation}
\alpha\wedge\star\alpha=(\alpha,\alpha)\star1\,,
\end{equation}
where $(\cdot,\cdot)$ is the scalar product associated to the metric $g$
introduced in the previous section. Moreover, the same scalar product can
be rewritten with the new convolution as
\begin{equation}
(\alpha,\alpha)=\alpha\bullet'\star\alpha=(-1)^{p\left(  n-p\right)  }%
\star\alpha\bullet'\alpha
\end{equation}
where instead of the wedge product we have used the convolution product.

\vskip 1cm

\noindent  {\bf Acknowledgements}

\noindent We thank Paolo Aschieri for valuable discussions.

\end{document}